\documentclass{ws-procs975x65}

\begin{document}

\title{Heavy quark chemical potential as probe of the phase diagram of nuclear matter}

\author{P. G. Katsas$^*$, A. D. Panagiotou $^{\$}$ and T. Gountras}

\address{University of Athens, Physics Department,\\
Nuclear and Particle Physics Division, \\ GR-15771 Athens,
Hellas\\ $^*$E-mail: pkatsas@phys.uoa.gr\\ $^{\$}$Email:
apanagio@phys.uoa.gr}

\maketitle

\abstracts{We study the temperature dependence of the strange and
charm quark chemical potentials in the phase diagram of nuclear
matter, within a modified and generalized hadron gas model, in
order to consider phase transitions and to describe phenomena
taking place outside the hadronic phase. We employ, in a
phenomenological way, the Polyakov loop and scalar quark
condensate order parameters, mass/temperature-scaled partition
functions and enforce flavor conservation. We propose that the
resulting variation of the heavy quark chemical potentials can be
directly related to the quark deconfinement and chiral phase
transitions. Then, the chemical potential of the strange and charm
quark can be considered as an experimentally accessible "order
parameter", probing the phase diagram of QCD.}

\section{Introduction}

One of the main problems in the study of the phase transitions,
occurring on the level of strong interactions, is finding an
unambiguous observable, which would act as an experimentally
accessible "order parameter" [1]. All proposed QGP signatures
(strangeness enhancement, J/$\psi$ suppression, dileptons,
resonance shift and broadening, etc.) have already been observed
in heavy ion collisions, however, we have seen, that they also
occur, to some extent, in $p-p$ or $p-A$ interactions where no QGP
production is theoretically expected. The physical quantity needed
should exhibit a uniform behavior within each phase, but should
change when a critical point is reached in a phase transition. It
has been earlier suggested [2-4] that the chemical potential of
strange quarks may be the sought-for macroscopic and therefore
measurable thermodynamic quantity. The case of [2+1] flavors was
thoroughly studied and it was shown that the change in the sign of
the strange quark chemical potential, from positive in the
hadronic phase to negative in the deconfined phase, may indeed be
a unique indication of the deconfinement phase transition. Here we
will review the basic aspects of the model and present the [2+2]
flavors version, which is a generalization of the model with the
inclusion of c-quark and charm hadrons.

\section{Hadronic phase}\label{sec:hg}
\begin{figure}[ht]
\centering
\epsfxsize=8cm  
\epsfbox{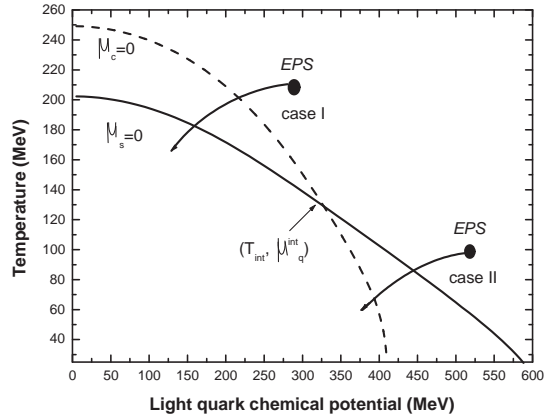} \caption{The critical curves $\mu_s=0$ and
$\mu_c=0$. We distinguish two cases depending on the location of
the equilibrated primordial state (EPS).} \label{fig:hgsc2}
\end{figure}

Assuming that the system has attained thermal and chemical
equilibration of four quark flavors $(u,d,s,c)$, the partition
function for the hadronic gas is written in the Boltzmann
approximation:
\begin{equation}
lnZ_{HG}(T,V,\lambda_s,\lambda_c)=lnZ_{HG}^{u,d}+lnZ_{HG}^{strange}+lnZ_{HG}^{charm}
\end{equation}
where
\begin{eqnarray}
lnZ_{HG}^{u,d}=Z_m+Z_n(\lambda_q^3+\lambda_q^{-3})\\
~~~~~~~~~~\nonumber\\
lnZ_{HG}^{strange}=Z_K(\lambda_q\lambda_s^{-1}+\lambda_q^{-1}\lambda_s)
+Z_Y(\lambda_q^2\lambda_s+\lambda_q^{-2}\lambda_s^{-1})\nonumber\\
+Z_\Xi(\lambda_q\lambda_s^2+\lambda_q^{-1}\lambda_s^{-2})
+Z_\Omega(\lambda_s^3+\lambda_s^{-3})
\end{eqnarray}
and
\begin{eqnarray}
lnZ_{HG}^{charm}=Z_D(\lambda_c\lambda_q^{-1}+\lambda_q\lambda_c^{-1})
+Z_{D_s}(\lambda_c\lambda_s^{-1}+\lambda_c^{-1}\lambda_s)\nonumber\\
+Z_{\Lambda_c,\Sigma_c}(\lambda_c\lambda_q^2+\lambda_c^{-1}\lambda_q^{-2})
+Z_{\Xi_c}(\lambda_q \lambda_s
\lambda_c+\lambda_q^{-1}\lambda_s^{-1}\lambda_c^{-1})\nonumber\\
+Z_{\Omega_c}(\lambda_s^2\lambda_c+\lambda_s^{-2}\lambda_c^{-1})
\end{eqnarray}
is the partition function for the non strange, strange and charm
sectors, respectively. The charm sector also includes
strange/charm mesons and baryons that lead to a coupling of the
fugacities $ \lambda_c, \lambda_s $. For simplicity we have
assumed isospin symmetry $\lambda_u=\lambda_d=\lambda_q$, while
the one particle Boltzmann partition function is given by:
\begin{equation}\label{bolap}
Z_k(V,T)=\frac{VT^3}{2\pi^2}\sum_jg_j\Big(\frac{m_j}{T}\Big)^2K_2\Big(\frac{m_j}{T}\Big)
\end{equation}
The summation in Eq.(\ref{bolap}) runs over the resonances of each
hadron species with mass $ m_j $, and the degeneracy factor $ g_j
$ counts the spin and isospin degrees of freedom of the
j-resonance. For the strange hadron sector, kaons with masses up
to 2045 MeV/$ c^2 $, hyperons up to 2350 MeV/$ c^2 $ and cascades
up to 2025 MeV/$ c^2 $ are included, as well as the $ \Omega^- $
states at 1672 MeV/$ c^2 $ and 2252 MeV/$ c^2 $. For the charm
hadron sector, we include purely charm mesons $D^+,D^-,D^0$ and
baryons ($\Lambda_c, \Sigma_c$) as well as strange-charm mesons
($D_s^{\pm}$) and baryons ($\Xi_c, \Omega_c$) which contain both
heavy quark flavors. All known charm resonances are taken into
account with masses up to 2.7 GeV/$ c^2 $. To derive the Equation
of State (EOS) of the hadron gas phase we simultaneously impose
flavor conservation,
\begin{eqnarray}
<N_s-N_{\overline{s}}> & = &
\frac{T}{V}\frac{\partial}{\partial\mu_s}lnZ_{HG}(T,V,\lambda_q,\lambda_s,\lambda_c)=0\\
<N_c-N_{\overline{c}}> & = &
\frac{T}{V}\frac{\partial}{\partial\mu_c}lnZ_{HG}(T,V,\lambda_q,\lambda_s,\lambda_c)=0
\end{eqnarray}
which reduce to a set of coupled equations:
\begin{eqnarray}\label{eq:hgs}
Z_K(\lambda_q^{-1}\lambda_s-\lambda_q\lambda_s^{-1})
+Z_Y(\lambda_q^2\lambda_s-\lambda_q^{-2}\lambda_s^{-1})+2Z_{\Xi}(\lambda_q\lambda_s^2-\lambda_q^{-1}\lambda_s^{-2})
\nonumber\\
+3Z_{\Omega}(\lambda_s^3-\lambda_s^{-3})+Z_{\Xi_c}(\lambda_q
\lambda_s
\lambda_c-\lambda_q^{-1}\lambda_s^{-1}\lambda_c^{-1})+2Z_{\Omega_c}
(\lambda_s^2\lambda_c-\lambda_s^{-2}\lambda_c^{-1})=0\nonumber\\~~~\\
Z_D(\lambda_c\lambda_q^{-1}-\lambda_q\lambda_c^{-1})
+Z_{D_s}(\lambda_c\lambda_s^{-1}-\lambda_c^{-1}\lambda_s)+
Z_{\Lambda_c,\Sigma_c}(\lambda_c\lambda_q^2-\lambda_c^{-1}\lambda_q^{-2})\nonumber\\
+Z_{\Xi_c}(\lambda_q \lambda_s
\lambda_c-\lambda_q^{-1}\lambda_s^{-1}\lambda_c^{-1})
+Z_{\Omega_c}(\lambda_s^2\lambda_c-\lambda_s^{-2}\lambda_c^{-1})=0~~~~~~~~~~
\end{eqnarray}
The above conditions, define the relation between all quark
fugacities and temperature in the equilibrated primordial state.
In the HG phase with finite net baryon number density, the
chemical potentials $\mu_q$, $\mu_s$ and $\mu_c$ are coupled
through the production of strange and charm hadrons. Due to this
coupling $\mu_s,\mu_c>0$ in the hadronic domain. A more elegant
formalism describing the HG phase is the Strangeness-including
Statistical Bootstrap model (SSBM) [5,6]. It includes the hadronic
interactions through the mass spectrum of all hadron species, in
contrast to other ideal hadron gas formalisms. The SSBM is
applicable only within the hadronic phase, defining the limits of
this phase. In the 3-flavor case, the hadronic boundary is given
by the projection on the 2-dimensional (T, $\mu_q$) phase diagram
of the intersection of the 3-dimensional bootstrap surface with
the strangeness-neutrality surface ($\mu_s$ = 0). Note that the
vanishing of $\mu_s$ on the HG borderline does not $apriori$
suggest that $\mu_s$ = 0 everywhere beyond. It only states that
the condition $\mu_s$ = 0 characterizes the end of the hadronic
phase. Figure~\ref{fig:hgsc2} exhibits the hadronic boundary for
two heavy quark flavors, obtained by imposing the conditions
$\mu_s=0$ and $\mu_c=0$ to Eq's.~(\ref{eq:hgs}),~(9). Observe,
that there exists an intersection point, at $T_{int}\sim130$ MeV
and $\mu_q^{int}\sim325$ MeV. For an equilibrated primordial state
(EPS) above this temperature, i.e $T>130$ MeV, and low $\mu_q$
values, we observe that as the temperature decreases, the
condition $\mu_c=0$ is realized before the vanishing of $\mu_s$
(case I), whereas for $T<130$ MeV and high $\mu_q$, the opposite
effect takes place (case II). This behavior, may be of some
importance towards a possible experimental identification of a
color superconducting phase, which is realized at a low
temperature and high density region of the phase diagram (case
II).

\section{Chirally symmetric QGP phase}
The partition function for a four flavor Quark Gluon plasma has
the form,
\begin{eqnarray}\label{qgp}
lnZ_{QGP}(T,V,\mu_{q,s,c}) &=&
\frac{V}{T}\Big[\frac{37}{90}\pi^2T^4+\mu_q^2T^2+\frac{\mu_q^4}{2\pi^2}\nonumber\\
&~&~~~~~~~~+\sum_{i=s,c} \frac{g_i m_i^{02}
T^2}{2\pi^2}(\lambda_i+\lambda_i^{-1})K_2\Big(\frac{m_i^0}{T}\Big)\Big]
\end{eqnarray}
where $m_s^0$, $m_c^0$ is the current strange and charm quark
masses respectively. Flavor conservation within the QGP phase
yields $\lambda_s=\lambda_c=1$ or
\begin{equation}
\mu_s^{QGP}(T,\mu_q,\mu_c)=\mu_c^{QGP}(T,\mu_q,\mu_s)=0
\end{equation}
throughout this phase. Here the two order parameters, the Polyakov
loop $<L>$ and the scalar quark density $<\overline{\psi}\psi>$ ,
have reached their asymptotic values. Note that the chirally
symmetric quark-gluon plasma phase always corresponds to a
vanishing heavy quark chemical potential.

\section{Deconfined Quark Matter phase of [2+2] flavors}

We argue that, beyond the hadronic phase, an intermediate domain
of deconfined yet massive and correlated quarks arises, according
to the following qualitative picture: The thermally and chemically
equilibrated primordial state at finite baryon number density,
consists of the deconfined valance quarks of the participant
nucleons, as well as of $q-\overline{q}$ pairs, created by quark
and gluon interactions. Beyond but near the HG boundary, T $\geq$
$T_d$, the correlation-interaction between $q-q$ is near maximum,
$\alpha_s(T)\leq$1, a prelude to confinement into hadrons upon
hadronization. With increasing temperature, the correlation of the
deconfined quarks gradually weakens, $\alpha_s(T)\rightarrow$ 0,
as color mobility increases. The mass of all (anti)quarks depends
on the temperature and scales according to a prescribed way. The
initially constituent mass decreases with increasing $T>T_d$, and
as the DQM region goes asymptotically into the chirally symmetric
QGP phase, as T $\rightarrow$ $T_\chi$, quarks attain current
mass. Thus, we expect the equation of state in the intermediate
DQM region to lead to the EoS of the hadronic phase, Eq. (1), at T
$\leq$ $T_d$, and to the EoS of the QGP, Eq. (6), at T $\sim$
$T_\chi$. In order to construct an empirical partition function
for the desciption of the DQM phase, we use (a) the Polyakov loop
$<L>\sim e^{-F_q/T}\equiv R_d(T \geq T_d)=0 \rightarrow 1$ as
T=$T_d \rightarrow T_\chi$ and (b) the scalar density
$<\overline{\psi}\psi> \equiv R_\chi(T \geq T_d)=1 \rightarrow 0$
as T=$T_d \rightarrow T_\chi$. The first describes the quark
deconfinement while the latter is associated with the quark mass
scaling.We assume that above the deconfinement temperature, quarks
retain some degree of correlation and can be considered as
hadron-like states. Therefore, near $T_d$ a hadronic formalism may
still be applicable. This correlation/interaction gradually
weakens, as a result of the progressive increase of color
mobility. Each quark mass scales, decreasing from the constituent
value to the current one as we reach the chiral symmetry
restoration temperature ($T\rightarrow T_\chi$). Thus, we consider
a temperature dependent mass for each quark flavor, approximated
by:
\begin{equation}\label{qmass}
m_f^*(T)=R_{\chi}(T)(m_f-m_f^0)+m_f^0
\end{equation}
where $m_f$ and $m_f^0$ are the constituent and current quark
masses respectively (the values $m_u^0=5MeV, m_d^0=9MeV,
m_s^0=170MeV, m_c^0=1.1GeV$ have been used). In the same spirit,
we approximate the effective hadron-like mass:
\begin{equation}\label{hadmass}
m_i^*(T)=R_{\chi}(T)(m_i-m_i^0)+m_i^0
\end{equation}
where $m_i$ is the mass of each hadron in the hadronic phase and
$m_i^0$ is equal to the sum of the hadron's quarks current mass
(for example $m_K^0=175MeV, m_{\Xi}^0=350MeV$). In the partition
function of the DQM phase, the former scaling is employed through
the mass-scaled QGP partition function $lnZ_{QGP}^*$, where all
quark mass terms are given by Eq.(\ref{qmass}), while the latter
is used in the mass-scaled hadronic partition function
$lnZ_{HG}^*$, where all hadron mass terms are given by
Eq.(\ref{hadmass}). Employing the described dynamics, we construct
an empirical partition function for the DQM phase,
\begin{eqnarray}\label{DQM}
lnZ_{DQM}(V,T,\{\lambda_f\}) &=& [1-R_d(T)]lnZ_{HG}^*(V,T,\{\lambda_f\})\nonumber\\
&~&~~~~~~+R_d(T)lnZ_{QGP}^*(V,T,\{\lambda_f\})~~~~~~
(f=q,s,c)\nonumber
\end{eqnarray}
The factor $[1-R_d(T)]$ describes the weakening of the interaction
of the deconfined quarks, while the factor $R_d(T)$ can be
associated with the increase of color mobility as we approach the
chirally symmetric QGP phase. The DQM partition function is a
linear combination of the HG and QGP mass-scaled partition
functions together with the general demand to describe both
confinement and chiral symmetry restoration asymptotically. Note
that below the deconfinement critical point $T<T_d$, $R_d(T)=0$,
leading to $lnZ_{DQM}=lnZ_{HG}$ (with constituent quarks), whereas
at the chiral symmetry restoration temperature $T \sim T_{\chi}$,
$R_d(T)=1$ and $lnZ_{DQM}=lnZ_{QGP}$ (with current quark masses).
In order to acquire the EoS of the DQM phase, we impose again the
strangeness and charm neutrality conditions, leading to the set of
equations respectively,
\begin{eqnarray}\label{fig:dqms}
[1-R_d(T)]\Big[Z_K^*(\lambda_s\lambda_q^{-1}-\lambda_q\lambda_s^{-1})+
Z_Y^*(\lambda_s\lambda_q^2-\lambda_s^{-1}\lambda_q^{-2})\nonumber\\
+2Z_{\Xi}^*(\lambda_s^2\lambda_q- \lambda_s^{-2}\lambda_q^{-1})
+3Z_{\Omega}^*(\lambda_s^3-\lambda_s^{-3})+Z_{\Xi_c}^*(\lambda_q
\lambda_s
\lambda_c-\lambda_q^{-1}\lambda_s^{-1}\lambda_c^{-1})\nonumber\\
+2Z_{\Omega_c}^*(\lambda_s^2\lambda_c-\lambda_s^{-2}\lambda_c^{-1})\Big]+R_d(T)g_s
m_s^{*2}K_2\Big(\frac{m_s^*}{T}\Big)(\lambda_s-\lambda_s^{-1})=0~~~~
\end{eqnarray}
and
\begin{eqnarray}\label{fig:dqmc}
[1-R_d(T)]\Big[Z_D^*(\lambda_c\lambda_q^{-1}-\lambda_q\lambda_c^{-1})
+Z_{D_s}^*(\lambda_c\lambda_s^{-1}-\lambda_c^{-1}\lambda_s)+\nonumber\\
Z_{\Lambda_c,\Sigma_c}^*(\lambda_c\lambda_q^2-\lambda_c^{-1}\lambda_q^{-2})
+Z_{\Xi_c}^*(\lambda_q \lambda_s
\lambda_c-\lambda_q^{-1}\lambda_s^{-1}\lambda_c^{-1})\nonumber\\
+Z_{\Omega_c}^*(\lambda_s^2\lambda_c-\lambda_s^{-2}\lambda_c^{-1})\Big]+
R_d(T)g_c
m_c^{*2}K_2\Big(\frac{m_c^*}{T}\Big)(\lambda_c-\lambda_c^{-1})=0
\end{eqnarray}
which must be solved simultaneously. Note that because of the
strange/charm hadrons $D_s, \Xi_c, \Omega_c$ there exists a
coupling between the heavy quark fugacities $\lambda_s,
\lambda_c$. By solving the above equations, for a given chemical
potential $\mu_q$, we derive the variation of the strange and
charm quark chemical potentials with temperature in the phase
diagram.

\section{Results for finite chemical potential}

\begin{figure}[hb]
\centering
\epsfxsize=8cm   
\epsfbox{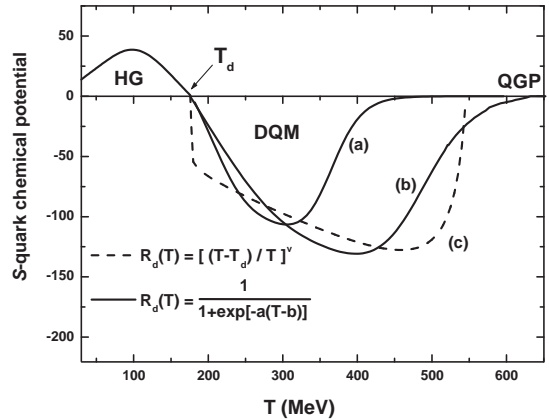} \caption{Variation of $\mu_s$ with the
temperature in the case of [2+1] quark flavors and different
approximations or parameterizations of the order parameter
$R_d(T)$.} \label{fig:1}
\end{figure}

\begin{figure}[ht]
\centering \epsfxsize=7cm \epsfbox{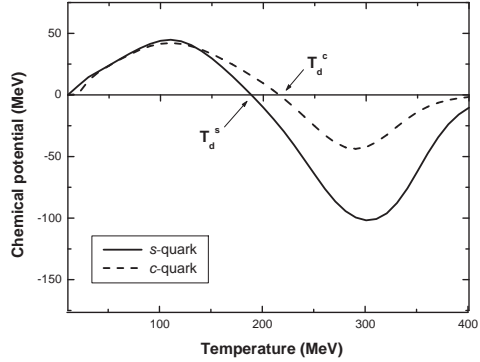} \caption{Plot of
the strange and charm quark chemical potentials in the phase
diagram for $\lambda_q=0.48$. Notice that the change in their sign
is realized at different temperatures.} \label{fig:mc}
\end{figure}

In the case of 3-flavors and finite density, we had neglected all
terms involving c-quarks ($\lambda_c=1$). In this case, only the
variation of $\mu_s$ was considered and Figure~\ref{fig:1} was
derived. We observe that the strange quark chemical potential
attains positive values in the hadronic phase, becomes zero upon
deconfinement, it grows strongly negative in the DQM domain and
finally returns to zero as the QGP phase is approached. It is
important that $\mu_s$ behaves differently in each phase, as this
is what we are looking for from the beginning in the search for an
experimentally accessible "order parameter". The change in the
sign of $\mu_s$ from positive in the hadronic phase to negative in
the deconfined is an unambiguous indication of the quark
deconfinement phase transition, as it is independent of
assumptions regarding interaction mechanisms. In the case of [2+2]
flavors the situation is slightly modified. Figure \ref{fig:mc}
exhibits the variation of the two correlated heavy quark chemical
potentials with the temperature of the primordial state, as given
by Eq's~(\ref{fig:dqms}),~(\ref{fig:dqmc}). We observe that both
are initially positive and then grow negative, although the change
in their sign is realized at different temperatures, for example
$\mu_s=0$ at $T_d^s\sim190 MeV$, while $\mu_c=0$ at $T_d^c\sim215$
MeV for a fixed value of the fugacity $\lambda_q=0.48$. However,
this difference can be easily understood if we consider
Figure~\ref{fig:mc} in the framework of Figure~\ref{fig:hgsc2}. As
already discussed in Sec.~\ref{sec:hg}, for an equilibrated
primordial state ($EPS$) with $T>T_{int}$ and sufficiently low
$\mu_q$, $\mu_c$ becomes zero earlier than $\mu_s$, as the system
approaches hadronization (see Figure~\ref{fig:hgsc2}).
This is the reason why $T_d^c>T_d^s$ in Figure~\ref{fig:mc}. For
sufficiently high $\mu_q$ values and low temperatures, the
opposite effect is present, i.e $\mu_c$ changes it's sign at a
lower temperature than the strange quark chemical potential. The
magnitude of the difference $|T_d^s-T_d^c|$, will depend on the
exact location of the state in the phase diagram. The fact that
the $\mu_s, \mu_c$ vanish at different temperatures, at the end of
the respective hadronic domain, has further consequences, as it
implies that there exists a quark "deconfinement region" rather
than a certain critical line.

\section{Experimental data}\label{sec:data}
\begin{figure}[ht]
\centering \epsfxsize=7 cm \epsfbox{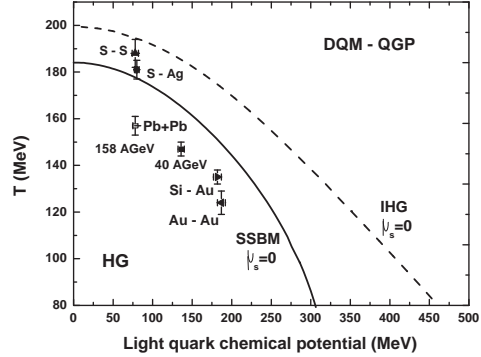} \caption{(T,
$\mu_q$) values of several interactions and their location in the
phase diagram. The lines correspond to the hadronic boundary
within the SSB and IHG models.} \label{fig:data}
\end{figure}

Over the last years, data from several nucleus-nucleus collisions
have been analyzed within thermal statistical models, employing
the canonical and grand-canonical formalisms [7-11]. Table 1
summarizes some of the results for the quantities T, $\mu_q$ and
$\mu_s$, which have been deduced after performing a fit to the
experimental data. Figure \ref{fig:data} shows the phase diagram
with the Ideal Hadron Gas (IHG) and SSBM $\mu_s=0$ lines, as well
as the location of the mean (T, $\mu_q$) values obtained for every
collision. We observe that all interactions studied, are
consistently situated inside the hadronic phase, defined by the
IHG model and exhibit positive $\mu_s$. The sulfur-induced
interactions, however are situated slightly beyond the hadronic
phase defined by the SSBM. IHG calculations exhibit deviations
from the SSBM as we approach the critical deconfinement point
$T=T_d \sim 175$ MeV, where the S-S and S-Ag interactions are
roughly located. Within the IHG model the condition $\mu_s=0$ is
satisfied at a higher temperature ($T \sim 200$ MeV), extending
the hadronic phase to a larger region as can be seen in Figure
\ref{fig:data}. As a consequence, $\mu_s$ changes sign at a higher
temperature also and this is the reason why $\mu_s>0$ in the
analysis of [11], although a temperature above deconfinement
(according to the SSBM) is obtained. Therefore, an adjustment of
the IHG curve to the SSBM boundary and a new fit to the data are
necessary [12]. The data from RHIC at $\sqrt{s}$=130, 200 AGeV are
not included in our discussion, since at such high energies
$\mu_q$ is very small and $\mu_s \sim 0$ throughout the phase
diagram. The observation of negative heavy quark chemical
potential requires a finite baryon density system.

\begin{table}[t]
  \centering
\small{
\begin{tabular}{lllll}

\multicolumn{5}{l}{\textbf{Table 1}. Deduced values for T,
$\mu_q$, $\mu_s$ from several thermal} \\
\multicolumn{5}{l}{models and fits to experimental data for
several interactions.}\\ \hline \hline
\multicolumn{5}{c}{\textbf{Interaction/Experiment}}\\ \hline
\multicolumn{5}{c}{Si+Au(14.6 AGeV)/E802}\\
   & Reference[4] & Reference[9] & Mean & \\
  T(MeV) & 134$\pm$6 & 135$\pm$4 & 135$\pm$3 & \\
  $\mu_q$(MeV) & 176$\pm$12 & 194$\pm$11 & 182$\pm$5 &  \\
  $\mu_s$(MeV) & 66$\pm$10 & & 66$\pm$10 & \\
  \multicolumn{5}{c}{Pb+Pb(158 AGeV)/NA49}\\
  & Reference[4] & Reference[9] & Reference[7] & Mean \\
  T(MeV) & 146$\pm$9 & 158$\pm$3 & 157$\pm$4 & 157$\pm$3\\
  $\mu_q$(MeV) & 74$\pm$6 & 79$\pm$4 & 81$\pm$7 & 78$\pm$3 \\
  $\mu_s$(MeV) & 22$\pm$3 & & 25$\pm$4 & 23$\pm$2\\
  \multicolumn{5}{c}{Pb+Pb(40 AGeV)/NA49}\\
  & Reference[4] & Reference[*] & Mean & \\
  T(MeV) & 147$\pm$3& 150$\pm$8 & 149$\pm$9 & \\
  $\mu_q$(MeV) & 136$\pm$4 & 132$\pm$7 & 134$\pm$8 & \\
  $\mu_s$(MeV) & 35$\pm$4 & & & \\
  \multicolumn{5}{c}{S+S(200 AGeV)/NA35}\\
  & Reference[10] & Reference[11] & Reference[8] & Mean \\
  T(MeV) & 182$\pm$9 & 181$\pm$11 & 202$\pm$13 & 188$\pm$6\\
  $\mu_q$(MeV) & 75$\pm$6 & 73$\pm$7 & 87$\pm$7 & 78$\pm$4 \\
  $\mu_s$(MeV) & 14$\pm$4 & 17$\pm$6 & & 16$\pm$7\\
  \multicolumn{5}{c}{S+Ag(200 AGeV)/NA35}\\
   & Reference[10] & Reference[11] & Reference[8] & Mean \\
  T(MeV) & 180$\pm$3 & 179$\pm$8 & 185$\pm$8 & 181$\pm$4\\
  $\mu_q$(MeV) & 79$\pm$4 & 81$\pm$6 & 81$\pm$7 & 80$\pm$3 \\
  $\mu_s$(MeV) & 14$\pm$4 & 16$\pm$5 & & 16$\pm$8\\
   & & & & \\
  \multicolumn{5}{l}{*NA49 private communication}\\
  \hline \hline
  \end{tabular}
}
 \end{table}

\section{Conclusions}

On the basis of the present analysis, we conclude that the heavy
quark chemical potentials behave differently in each region
(HG-DQM-QGP) of the phase diagram and, therefore, they can serve
as a probe of the phase transitions. This is the first proposal of
such an experimentally accessible "order parameter" that holds for
a finite baryon density state. The appearance of negative values
of $\mu_s$ and $\mu_c$, is a well-defined indication of the quark
deconfinement phase transition, at T=$T_d$, which is free of
ambiguities related to microscopic effects of the interactions. It
is important to add, that the observation of negative heavy quark
chemical potentials would be also a clear evidence for the
existence of the proposed DQM phase, meaning that chiral symmetry
and deconfinement are apart at finite density. Until now, there is
no known argument from QCD that the two transitions actually occur
at the same temperature. Au+Au collisions at intermediate
energies, for example $30 \leq \sqrt{s} \leq 90~AGeV$, should be
performed to experimentally test our proposals.

\section*{Acknowledgments}
P. Katsas is grateful to the organizing committee, for the
opportunity to participate in the conference. This work was
supported in part by the Research Secretariat of the University of
Athens.

\end{document}